\documentclass[rapids]{jfm}

\usepackage{graphicx,float,calc,ifthen}
\usepackage[fleqn]{amsmath}
\usepackage{quoting}
\usepackage{cancel}
\usepackage{amsmath,bm} 
\usepackage{amssymb}    
\usepackage{hyperref} 

\newcommand{\lmm}{\textcolor{black}}    
\newcommand{\lmr}{\textcolor{black}}
\newcommand{\lmf}{\textcolor{black}}
\newcommand{\lmb}{\textcolor{black}}

    \hypersetup{
      final=true,
      plainpages=false,
      pdfstartview=FitV,
      pdftoolbar=true,
      pdfmenubar=true,
      bookmarksopen=true,
      bookmarksnumbered=true,
      breaklinks=true,
      linktocpage,
      colorlinks=true,
      linkcolor=blue,
      urlcolor=blue,
      citecolor=blue,
      anchorcolor=green
      pagebackref
    }
\usepackage{lipsum}
\usepackage{natbib}
\usepackage{mathtools}

\shorttitle{Sensitivity of the Rayleigh criterion in thermoacoustics}
\shortauthor{Luca Magri, Matthew P. Juniper and Jonas P. Moeck}

\title{Sensitivity of the Rayleigh criterion in thermoacoustics}

\author{Luca Magri\aff{1}\corresp{\email{lm547@cam.ac.uk}}, Matthew P. Juniper\aff{1} \and Jonas P. Moeck\aff{2}
  }

\affiliation{\aff{1}Cambridge University Engineering Dept., Trumpington St, CB2 1PZ, Cambridge, UK
\aff{2}Department of Energy and Process Engineering, NTNU, Trondheim, Norway}

\begin{document}

\maketitle

\begin{abstract}
Thermoacoustic instabilities are one of the most challenging problems faced by gas turbine and rocket motor manufacturers. 
The key instability mechanism is described by the {\it Rayleigh criterion}. 
The Rayleigh criterion does not directly show how to alter a system to make it more stable. This is the objective of sensitivity analysis. 
Because thermoacoustic systems have many design parameters, adjoint sensitivity analysis has been proposed to obtain  all the sensitivities with one extra calculation. 
Although adjoint sensitivity analysis can be carried out in both the time and the frequency domain, the frequency domain is more natural for a linear analysis. 
Perhaps surprisingly, the Rayleigh criterion has not yet been rigorously derived and comprehensively interpreted in the frequency domain. 
The contribution of this theoretical paper is threefold.  
First, the Rayleigh criterion is interpreted in the frequency domain with integral formulae for the complex eigenvalue. 
Second, the first variation of the Rayleigh criterion is calculated both in the time and frequency domain, both with and without Lagrange multipliers (adjoint variables). 
The Lagrange multipliers are physically related to the system's observables. 
Third, an adjoint Rayleigh criterion is proposed.
The paper also points out that the conclusions of {\it Juniper, M. P. (2018),  Phys. Rev. Fluids, vol. 3, 110509} apply to the first variation of the Rayleigh criterion, not to the Rayleigh criterion itself. 
The mathematical relations of this paper can be used to compute sensitivities directly from measurable quantities to enable optimal design.  
    \end{abstract}

   %
\section{Introduction}
The most commonly used criterion to assess the stability of a thermoacoustic system, such as a gas turbine combustor or a rocket motor, is the Rayleigh criterion~\citep{Rayleigh1878}. 
This criterion states that the acoustic energy increases when the heat release rate is sufficiently in phase with the acoustic pressure that the work done by gas expansion exceeds the energy loss over a cycle.
The Rayleigh criterion has been used in experiments, numerical simulations, and theory since the 1940's~\citep{Juniper2018}. 
The literature is thoroughly reviewed by~\citet{Candel2002,Lieuwen2005,Poinsot2017,Juniper2018} and here we describe only a few examples.  
In experiments~\cite[e.g.,][]{Lieuwen2005}, the Rayleigh criterion is used as a diagnostic tool. The spatial product between the acoustic pressure and the heat release rate, $p\dot{Q}$, also known as the Rayleigh index, is used to identify spatial locations where the thermoacoustic instability is amplified or damped. 
In active control of thermoacoustic oscillations~\cite[e.g.,][]{Dowling2005}, the Rayleigh index is used to measure the influence of the heat release caused by a secondary fuel injection; if the Rayleigh index becomes negative, the secondary fuel damps out the instability.
In large eddy simulations~\cite[e.g.,][]{Poinsot2017}, the Rayleigh criterion has been used to analyse the thermoacoustic instability of staged and annular combustors. 
In low-order models of annular combustors, the Rayleigh criterion provides conditions for the stability of spinning and standing waves~\citep{Ghirardo2016}. 
\lmr{The Rayleigh criterion is the most widely used tool in thermoacoustic stability, but it does not take into account the effect of a mean flow. Other more general criteria have been developed to account for a mean flow~\citep{Chu1965,Candel1975,Myers1991,Nicoud2005,Karimi2008,JosephGeorge2011,George2012} and multi-component gases~\citep{Brear2012}. In this paper, we focus on the Rayleigh criterion.}

Although the Rayleigh index indicates the regions that contribute to the instability, it does not directly provide sensitivity information on how to control the instability.
To address the design question ``where and how should the thermoacoustic system be changed in order to be optimally stabilized'', adjoint sensitivity analysis was developed for optimal passive control. 
This is reviewed by~\cite{Magri2019_amr} and tutorials can be found in \cite{Juniper2018_prf}. \lmr{In non-reacting flows, adjoint analysis is reviewed by~\citet{Sipp2010,Luchini2014,Camarri2015}.  }
In sensitivity analysis in the frequency domain, \cite{Magri2013} compared the Rayleigh index with eigenvalue sensitivities for passive control of the oscillations in a time-lag model, and \cite{Magri2013c} compared the Rayleigh index of a diffusion flame with eigenvalue sensitivity to identify the most sensitive regions of the flame.  
The Rayleigh criterion was useful to gain insight into the active physical mechanisms leading to instability, while sensitivity analysis was useful to make optimal changes to the system.

\citet[\S III, ][]{Chu1965} examines the behaviour of acoustic disturbances subject to arbitrary injections of mass, momentum, and heat.
He asks the question ``what is the contribution of an arbitrary injection of heat to the energy of an acoustic oscillation" and shows that the rate of change of energy is the integral of the injected heat release rate multiplied by the \lmr{temperature} (with adjustments for dissipation, \citet[eq. 19a,][]{Chu1965}). \lmr{If the thermal conductivity is zero}, the rate of change of energy becomes the integral of the injected heat release rate multiplied by the \lmr{pressure} (with adjustments for dissipation, \citet[eq. 23a,][]{Chu1965}). 
This is the Rayleigh criterion. 
%
%
On the other hand, sensitivity analysis such as \cite{Magri2013,Magri2013c,Juniper2018_prf} examines the behaviour of thermoacoustic disturbances subject to infinitesimal injections of mass, momentum, and heat. 
They ask the question ``what is the contribution of a small injection of heat to the energy of a \emph{thermo}acoustic oscillation" and show that the rate of change of energy is the integral of the injected heat release rate multiplied by the \emph{adjoint} pressure. 
This is best described as the first variation of the Rayleigh criterion.
The Rayleigh criterion is formed with the physical pressure, while the first variation of the Rayleigh criterion is formed with the adjoint pressure.
The conclusions of \cite{Juniper2018_prf} should have been more carefully worded with regard to this point.

In this paper, we first derive and interpret the Rayleigh criterion. The novelty is the analysis in the frequency domain.
Second, we derive and interpret the first variation of the Rayleigh criterion with and without Lagrange multipliers (adjoint variables).
The functional derivatives are calculated and interpreted as the sensitivities of the Rayleigh criterion.
Crucially, we physically interpret the adjoint variables in terms of observable quantities. 
Third, an adjoint Rayleigh criterion is proposed to interpret the sensitivity of thermoacoustic stability to perturbations to the heat source.  
%
\section{Time domain analysis}\label{sec:timedomain}
\subsection{The Rayleigh criterion}
We consider a duct of length $L$ with a zero-Mach-number base flow and ideal acoustic boundary conditions, i.e., the acoustics do no work on the surroundings. The cut-off frequency is sufficiently high that the acoustics are longitudinal in the direction $x$. 
The dimensional equations, which are derived by linearizing the Euler and energy equations for ideal gases, are  
\begin{align} \label{eq:fjirgnvir}
{\bar{\rho}}\frac{\partial u}{\partial t} + \frac{\partial p}{\partial x} &= 0, \\ 
\frac{1}{\gamma\bar{p}}\frac{\partial p}{\partial t} + \frac{\partial u}{\partial x} &= \frac{(\gamma-1)}{\gamma\bar{p}}{\dot{Q}},  \label{eq:fjirgnvir2}
\end{align}
where $\gamma$ is the heat-capacity ratio; 
$\bar{\rho}$ is the base-flow density, which can be a function of $x$; and  %
${\dot{Q}}$ is the heat release rate from a source. \lmr{To keep the theory as general as possible, we assume that $\dot{Q}$ is a prescribed function of time and space, i.e., 
$\dot{Q} = \dot{Q}(x,t)$. This function can consist of both open-loop terms, which are not functions of the acoustic variables, and closed-loop terms, which are functions of the acoustic variables. The closed-loop forcing term can be obtained from flame transfer functions  \citep[e.g.,][]{Lieuwen2005}. 
} 

The base-flow pressure, $\bar{p}$, is uniform and constant because of the zero-Mach-number assumption. Equations~\eqref{eq:fjirgnvir}-\eqref{eq:fjirgnvir2} define a model of a prototypical thermoacoustic system with a generic heat-source term. 
%
The operation $\int^T_0\int_0^L(u\cdot\eqref{eq:fjirgnvir}+p\cdot\eqref{eq:fjirgnvir2})\;dx\;dt$, where $[0, T]$ is an arbitrary interval of time, yields the governing equation for the change in the acoustic energy 
%
%
\begin{equation}
E \coloneqq \frac{1}{2}\left[\int_0^L\left(\bar{\rho}u^2 + \frac{p^2}{\gamma\bar{p}}\right)\;dx\right]^T_0 =\frac{\gamma-1}{\gamma\bar{p}}\int_0^T\int_0^L{p{\dot{Q}}}\;dx\;dt,   \label{eq:fflldd}
\end{equation}
where we considered that $up=0$ at $x=0$ and $x=L$ because the boundary conditions are ideal. The symbol $\coloneqq$ defines a quantity. 
For brevity, we refer to $E$ as the acoustic energy.
Equation~\eqref{eq:fflldd}, which has an arbitrary $T$, can be interpreted as an extension of the Rayleigh criterion: If the heat release rate of a source, ${\dot{Q}}$, is in phase with the acoustic pressure, $p$, on average  then the  acoustic energy of the duct, $E$, increases in time on average. Although strictly speaking the original Rayleigh criterion \citep{Rayleigh1878} takes $T$ to be the period of an oscillation, we refer to \eqref{eq:fflldd} as the Rayleigh criterion and to the integrand $p\dot{Q}$ as the Rayleigh index. 
%
%
%

We work in Hilbert spaces with the following inner products
\begin{align}
    \left\langle a, b\right\rangle_V \coloneqq \int_0^La^*b\;dx, \label{eq:innerproductddd}\quad\quad\quad
        \left\langle a, b\right\rangle_{V,T} \coloneqq \int_0^T\left\langle a, b\right\rangle_V\;dt, 
\end{align}
where ${}^*$ denotes the complex conjugate, and $a$ and $b$ are generic functions. 
In this section the variables are real, so the complex conjugate has no effect. 
We will, however, use the complex conjugate in the frequency domain (\S\ref{sec:fredomainanafff}). 
The acoustic energy is defined by a functional of two functions
\begin{align}
 E[p, \dot{Q}] \coloneqq  \left(\frac{\gamma-1}{\gamma\bar{p}}\right)\left\langle p, \dot{Q} \right\rangle_{V,T}, 
\end{align}
which provides a mapping from the acoustic pressure and heat release rate function spaces to a real number. We regard the heat release rate as the independent parameter. 
Physically, the acoustic pressure is a function of the heat release rate through the constraints imposed by the momentum and energy equations~\eqref{eq:fjirgnvir}-\eqref{eq:fjirgnvir2}. 
%

%

%
%
\subsection{First variation without Lagrange multipliers}\label{sec:firstvariation_time}

We are interested in calculating the effect that a variation to the heat release rate $
\dot{Q}(x,t)\rightarrow \dot{Q}(x,t) + \delta\dot{Q}(x,t)$ 
has on the acoustic energy. In particular, we wish to calculate the effect of an {\it arbitrary} small variation 
$\delta\dot{Q}(x,t) =\epsilon \dot{Q}_p(x,t)$, 
 where $\epsilon$ is infinitesimal and $\lmb{\dot{Q}_p}$ is an arbitrary function, known as the test function. (For clarity, the dependence on $(x,t)$ will be used only in some passages.) 
 The symbol $\delta$ refers to an infinitesimal virtual change, whereas the differentiation symbols $d$ or $\partial$ refer to an infinitesimal actual change.
The objective is to find the first variation of the acoustic energy, $E$, in the vicinity of $p$ and $\dot{Q}$
%
for any arbitrary perturbation, $\epsilon\dot{Q}_p$, to the heat release rate 
%
%
%
\begin{align}
{\delta E} = \epsilon  \int_0^T\int_0^L
\left(
\frac{\delta_{\dot{Q}} E}{\delta\dot{Q}(x',t')}+ 
    \frac{\delta_p E}{\delta p(x',t')}\frac{d p}{d \dot{Q}(x',t')}\right) \dot{Q}_p(x',t') \;dx'\; dt',    \label{eq:fijr3jfi11111}
\end{align}
 where the term within round brackets is the functional derivative, which is the kernel of the integral~\citep[e.g., ][]{Reinhardt1996}. 
 %
 %
The functional derivative of the acoustic energy consists of two terms. The first term is the sensitivity of the acoustic energy to an arbitrary perturbation to the heat release rate, $\delta_{\dot{Q}} E/\delta\dot{Q}(x',t')$. The second term  is the sensitivity to an arbitrary perturbation to the acoustic pressure, $\delta_p E/\delta p(x',t')$. The acoustic pressure, however, cannot arbitrarily change because it is a function of the heat release rate. The term $dp/d\dot{Q}$ is the sensitivity of the acoustic pressure to the virtual perturbation to the heat release rate 
\lmr{
\begin{align}
\frac{dp}{d\dot{Q}} \coloneqq \lim_{\epsilon\rightarrow0}\frac{p(\dot{Q}+\epsilon\dot{Q}_p) - p(\dot{Q})}{\epsilon\dot{Q}_p}, 
\end{align}
where $p(\cdot)$ signifies that the pressure is the solution of \eqref{eq:fjirgnvir}-\eqref{eq:fjirgnvir2} when the heat-release rate is provided by the argument $(\cdot)$. 
}
This constrains the pressure variation to be in the admissible function space imposed by the momentum and energy equations. 
Working on the assumption that the acoustic energy is Fr\'echet differentiable and taking the first variation of~\eqref{eq:fflldd} yields the two functional derivatives
\begin{align}
\frac{\delta_{\dot{Q}} E}{\delta\dot{Q}} & = \lmf{\left(\frac{\gamma-1}{\gamma\bar{p}}\right)}p, \quad\quad\quad\quad\quad\quad \frac{\delta_{p} E}{\delta p}  = \lmf{\left(\frac{\gamma-1}{\gamma\bar{p}}\right)}\dot{Q} , 
\end{align}
whence the functional derivative of the acoustic energy is 
\begin{align}
\frac{\delta E}{\delta\dot{Q}} \coloneqq \left(\frac{\delta_{\dot{Q}} E}{\delta\dot{Q}(x',t')}+ 
    \frac{\delta_p E}{\delta p(x',t')}\frac{d p}{d \dot{Q}(x',t')}\right) =  \left(\frac{\gamma-1}{\gamma\bar{p}}\right)\left(p+\frac{dp}{d\dot{Q}}\dot{Q}\right). \label{eq:fjfjfjfjfjfjw111}
\end{align}
Therefore, the first variation is   
$
\delta E =
(\gamma-1)/(\gamma\bar{p}) \langle p+({dp}/{d\dot{Q}})\dot{Q}, \epsilon \dot{Q}_p \rangle_{V,T} $. 
%
%
The functional derivative of the acoustic energy~\eqref{eq:fjfjfjfjfjfjw111}  is the superposition of two physical effects. 
The first is a purely acoustic effect in the absence of a main heat source, i.e. $\dot{Q}=0$, in which the derivative is the acoustic pressure. The maximum change in the acoustic energy is obtained when the heat release rate perturbation is applied in phase with the pressure and where the acoustic pressure is maximum. 
The second is a purely thermal effect when a main heat source is present, i.e., $\dot{Q}\neq 0$, in which the derivative of the acoustic energy is the acoustic pressure sensitivity weighted by the unperturbed heat release rate $\dot{Q}$. 
The maximum change in the acoustic energy is obtained when the heat release rate perturbation is such that the change in the acoustic pressure at the main source is maximum.  
The thermoacoustic case is the combination of the purely acoustic and thermal effects. To optimally perturb the acoustic energy we should optimize both the acoustic and thermal effects: the optimal solution is not strikingly apparent. It will become apparent in the next section in which the Lagrange multipliers (adjoint variables) are invoked.  
\subsection{First variation with Lagrange multipliers}\label{sec:firstvariation_time2}
 The adjoint variables can be interpreted as the Lagrange multipliers of a constrained optimization problem~\cite[e.g,][]{Sipp2010,Magri2019_amr}.  
This is why we interchangeably use the wording ``adjoint variables" and ``Lagrange multipliers". 
The objective of this section is the same as that of \S\ref{sec:firstvariation_time} but the first variation of the acoustic energy is expressed as a functional of the Lagrangian multiplier. 
The problem is cast as a constrained optimization problem: we wish to find the functional derivative ${\delta E}/{\lmm{\delta\dot{{Q}}}}$ subject to \eqref{eq:fjirgnvir}, \eqref{eq:fjirgnvir2}. 
%
%
%
%
%
%
%
We define a Lagrangian 
\begin{align}\label{eq:lagrangian} 
\mathcal{L}\{u, p, u^+, p^+, \dot{Q}\} &\coloneqq E- \left\langle u^+,  \eqref{eq:fjirgnvir} \right\rangle_{V,T} 
- \left\langle p^+, \eqref{eq:fjirgnvir2} \right\rangle_{V,T}, 
\end{align}
%
and take the first variation with respect to virtual changes of its arguments. On integration by parts, we obtain 
%
%
%
\begin{align}
\delta \mathcal{L}
 &= \left\langle\frac{\delta_u E}{\delta u}, \delta u\right\rangle_{V,T}
 +\left\langle\frac{\delta_p E}{\delta p}, \delta p\right\rangle_{V,T} -L\bar{\rho}[u^+\delta u]_0^T - L\frac{1}{\gamma\bar{p}}[p^+\delta p]_0^T\ldots \nonumber\\
&- T[u^+\delta p]_0^L
 - T[p^+\delta u]_0^L + 
\left\langle \bar{\rho}\frac{\partial u^+}{\partial t} + \frac{\partial p^+}{\partial x}+ \frac{(\gamma-1)}{\gamma\bar{p}}\frac{\lmm{\partial\dot{Q}}}{\partial u}p^+, \delta u\right\rangle_{V,T} \ldots \nonumber\\
&
+ 
\left\langle \frac{1}{\gamma\bar{p}}\frac{\partial p^+}{\partial t} + \frac{\partial u^+}{\partial x} + \frac{(\gamma-1)}{\gamma\bar{p}}\frac{\lmm{\partial\dot{Q}}}{\partial p}p^+,  \delta p\right\rangle_{V,T} +  \epsilon\left(\frac{\gamma-1}{\gamma\bar{p}}\right)\left\langle p^+,   \lmb{\dot{{Q}}_p}\right\rangle_{V,T}. \label{eq:gorgo4fjifeijfie}
\end{align}
%
From \eqref{eq:fjirgnvir}, \eqref{eq:fjirgnvir2} and \eqref{eq:lagrangian}, it becomes apparent that $\delta\mathcal{L}=\delta E$.
To eliminate all except the last term in~\eqref{eq:gorgo4fjifeijfie}, we define the adjoint problem   
\begin{align}
\bar{\rho}\frac{\partial u^+}{\partial t} + \frac{\partial p^+}{\partial x}+ \frac{(\gamma-1)}{\gamma\bar{p}}\frac{\lmm{\partial \dot{Q}}}{\partial u}p^+ &=0, \label{eq:Iamverytired}\\
\frac{1}{\gamma\bar{p}}\frac{\partial p^+}{\partial t} + \frac{\partial u^+}{\partial x}+ \frac{(\gamma-1)}{\gamma\bar{p}}\frac{\lmm{\partial\dot{Q}}}{\partial p}p^+&=0.\label{eq:Iamverytired2}
\end{align}
The adjoint initial and boundary conditions are defined such that they eliminate the boundary terms in~\eqref{eq:gorgo4fjifeijfie}. The partial derivatives $\partial\dot{Q}/\partial u$ and $\partial\dot{Q}/\partial p$ are \lmr{zero if the heat-release rate has no closed-loop terms. Otherwise, these derivatives can be obtained by differentiating the time-domain transformation of the flame transfer functions.}  
With the Lagrange multiplier, the first variation of the acoustic energy is simply 
${\delta E}  = ({\gamma-1})/({\gamma\bar{p}})\langle p^+, \epsilon\lmb{\dot{Q}_p}\rangle_{V,T}$;
hence the functional derivative is
 \begin{align}\label{eq:aaqe43}
\frac{\delta E}{\delta\dot{Q}} & = \left(\frac{\gamma-1}{\gamma\bar{p}}\right) p^+.  
 \end{align}
 The Lagrange multiplier, $p^+$ (multiplied by $({\gamma-1})/({\gamma\bar{p}})$), is the functional derivative of the acoustic energy. 
 Note that the calculation of $dp/d\dot{Q}$ in~\eqref{eq:fjfjfjfjfjfjw111} is computationally expensive because it is the derivative of the pressure with respect to an arbitrary perturbation to the heat release rate. 
%
By solving for the adjoint problem~\eqref{eq:Iamverytired}-\eqref{eq:Iamverytired2}, we obtain \lmr{directly} this sensitivity information.
\subsection{Relation between first variations with and without Lagrange multipliers}\label{sec:fkfkfkfkfk2111116}
The connection between \eqref{eq:fjfjfjfjfjfjw111} and \eqref{eq:aaqe43}, which is guaranteed by the Riesz representation theorem, is   
%
\begin{equation}
p^+ =  p + \frac{d p}{\lmm{d\dot{Q}}}\lmm{\dot{Q}.\label{eq:ffe3344d}}\end{equation}
This is a key equation in the time domain. It is the link between the adjoint pressure and the observable quantities, which gives physical meaning to the adjoint pressure.
Relation~\eqref{eq:ffe3344d} shows how to express the first variation of the Rayleigh criterion in the time domain in a non-self-adjoint system ($\dot{Q}\neq0$);  non-self-adjointness manifests itself via the thermal sensitivity $(d p/d\dot{Q})\dot{Q}$, which is a nonlocal effect. 
%
If the heat sources are localized as 
$\lmm{\dot{Q}(x,t)}   = \lmm{Q_f(t)\delta(x-x_f)}$ and $\lmm{\dot{Q}_p(x,t)} = \lmm{Q_p(t)\delta(x-x_p)}$, where $\delta(x-x_{f,p})$  is the Dirac delta centred at $x_{f,p}$, the adjoint pressure is $
p_p^+ =  p_p + {dp_f}\lmm{Q}_f/({\epsilon\lmm{{Q}_p}})$. \lmr{(A short derivation can be found in \S3 of the supplementary material.)} The subscripts $f$ and $p$ signify that the variable is evaluated at $x_f$ and $x_p$, respectively. 
%
%
%
%
To achieve the maximum first variation in the acoustic energy, the phase between ${Q_p}$ and the unperturbed pressure $p_p$ is such that $p_f$ shifts to be more in phase with $\lmm{Q_f}$.  
Using the Lagrange multiplier instead of the observables offers a more compact interpretation of the first variation of the Rayleigh criterion. 
%
To achieve the maximum first variation in the acoustic energy, the heat source, ${Q}_f$, should be perturbed at the location $x_p$ where the adjoint pressure, $p^+$, is maximum. 

\section{Frequency domain analysis}\label{sec:fredomainanafff}
To switch to the frequency domain, we use modal decomposition    
$
u(x,t) = \hat{u}(x)\exp(\sigma t)+\textrm{c.c.}$ and $
p(x,t)  = \hat{p}(x)\exp(\sigma t)+\textrm{c.c.}$ for the acoustic velocity and pressure, respectively.   ``c.c." stands for complex conjugate of the preceding term and $\hat{u}(x)$, $\hat{p}(x)$ and $\sigma$ are complex. The real part of $\sigma$, denoted $\sigma_r$, is the growth rate and the imaginary part, denoted $\sigma_i$, is the angular frequency. 
By substituting the modal decomposition into the governing equations~\eqref{eq:fjirgnvir}-\eqref{eq:fjirgnvir2}, 
we obtain  
\begin{align}\label{eq:lop43}
\bar{\rho}\sigma\hat{u}+ \frac{\partial\hat{p}}{\partial x} &= 0, \\
\frac{1}{\gamma\bar{p}}\sigma\hat{p}+ \frac{\partial\hat{u}}{\partial x} &= \frac{\gamma-1}{\gamma\bar{p}}\lmm{\hat{\dot{Q}}}.  
\label{eq:lop432}
\end{align}
\lmr{To keep the theory as general as possible, we assume that $\hat{\dot{Q}}$ is a prescribed spatial function, i.e., 
$\hat{\dot{Q}} = \hat{\dot{Q}}(x)$, which is in a linear closed loop with $\hat{p}$ and $\hat{u}$ (but usually it is a nonlinear function of~$\sigma$). Typically, the heat-release rate is specified by flame transfer functions  \citep[e.g.,][]{Lieuwen2005}. }
Equations~\eqref{eq:lop43}-\eqref{eq:lop432} define an eigenvalue problem where $\hat{u}$ and $\hat{p}$ are the acoustic velocity and pressure eigenfunctions associated with the eigenvalue $\sigma$. These quantities characterize the natural response of the thermoacoustic system. 
\subsection{The Rayleigh criterion}
In the frequency domain, the complex instantaneous acoustic energy varies as $E_t=(1/2)\exp(2\sigma t)\int_0^L(\bar{\rho}\hat{u}^2+\hat{p}^2/({\gamma\bar{p}}))\;dx$.
The eigenvalue $\sigma$ provides half the growth rate and angular frequency of the complex acoustic energy. As a consequence, interpreting the Rayleigh criterion in the frequency domain is equivalent to interpreting the eigenvalue. We achieve this by deriving three integral formulae for the eigenvalue in \S\ref{eq:lldafio30}. 
%
The acoustic energy is obtained by integrating $\partial E_{t}/\partial t$ over an arbitrary interval $[0, T]$. 
We set  %
$C\coloneqq \int_0^L(\bar{\rho}\hat{u}^2+\hat{p}^2/({\gamma\bar{p}}))\;dx$ 
as a constant normalization factor. It is straightforward to show that $C$ is a nontrivial complex number in a thermoacoustic system (\S1 of the supplementary material).  
Therefore, the first variation of the acoustic energy with respect to a virtual change in the eigenvalue is $\delta E = ({C}/{2})\delta\left(\exp(2\sigma T) -1\right)
%
%
 =  CT 
\delta\sigma$.
%
The calculation of the first variation of the acoustic energy is one-to-one related to the first variation of the eigenvalue, $\delta\sigma$. In light of this, we focus only on the calculation of $\delta\sigma$ in \S\S\ref{sec:ffflkpp000003}-\ref{eq:FirstvariationwithLagrangemultipliers}. 
%
%
%
%
%
\subsubsection{Eigenvalue integral formulae}\label{eq:lldafio30}
Performing the operation~$\langle\hat{u}^*,\eqref{eq:lop43}\rangle_V + \langle\hat{p}^*, \eqref{eq:lop432}\rangle_V$ and dividing by $C$
%
provides an  eigenvalue  functional
%
%
%
\begin{align}
\sigma[\hat{p}^*, \hat{\dot{Q}}] \coloneqq \frac{1}{C}\left(\frac{\gamma-1}{\gamma\bar{p}}\right)\left\langle \hat{p}^*, \hat{\dot{Q}} \right\rangle_V,  \label{eq:ppooiill893}
\end{align}
which was used in~\cite{Magri2013} without any derivation. 
This relation is needed to define the eigenvalue functional derivative in \S\ref{sec:ffflkpp000003}. It is, however, difficult to interpret~\eqref{eq:ppooiill893} physically owing to the complex normalization factor $C$, which dephases the eigenvalue with respect to the numerator.  To enable physical interpretation of the Rayleigh criterion in the frequency domain, we derive two other integral formulae with real denominators, which do not affect the phase. 
Performing the operation~$Re$($\langle\hat{u},\eqref{eq:lop43}\rangle_V + \langle\hat{p}, \eqref{eq:lop432}\rangle_V$), 
after including the fact that the real part of $\int_0^L(\hat{u}^*\frac{\partial\hat{p}}{\partial x}
 + \hat{p}^*\frac{\partial\hat{u}}{\partial x})\;dx$ is zero (but the imaginary part is not) because of the ideal boundary conditions, offers a convenient functional  for the growth rate 
\begin{equation}\label{eq:ffrwu52478}
\sigma_r[\hat{p}, \hat{\dot{Q}}] = \frac{1}{F}
\left(\frac{\gamma-1}{\gamma\bar{p}}\right)
Re\left\langle\hat{p}, \hat{\dot{Q}}\right\rangle_V,  
\end{equation}
where $F\coloneqq {\int_0^L\left(\bar{\rho}|\hat{u}|^2+|\hat{p}|^2/({\gamma\bar{p}})\right)dx}$. 
%
%
%
%
%
%
%
%
On the other hand, for the angular frequency, performing the operation~$Im$($\langle\hat{u},\eqref{eq:lop43}\rangle_V - \langle\hat{p}, \eqref{eq:lop432}\rangle_V$), 
after including the fact that the imaginary part of $\int_0^L(\hat{u}^*\frac{\partial\hat{p}}{\partial x}
 - \hat{p}^*\frac{\partial\hat{u}}{\partial x})\;dx$ is zero (but the real part is not) because of the ideal boundary conditions, offers a convenient functional for the angular frequency  
\begin{equation}\label{eq:llqqkgj4893}
\sigma_i[\hat{p}, \hat{\dot{Q}}] = -\frac{1}{G} \left(\frac{\gamma-1}{\gamma\bar{p}}\right)
Im\left\langle\hat{p}, \hat{\dot{Q}}\right\rangle_V, 
\end{equation}
where
$G\coloneqq {\int_0^L(\bar{\rho}|\hat{u}|^2 - |\hat{p}|^2/({\gamma\bar{p}}))\;dx}\neq0$ (when $\hat{\dot{Q}}\neq0)$ following the argument of \S1 of the supplementary material.  When the heat release rate tends to zero, $\eqref{eq:llqqkgj4893}$ becomes indeterminate ($0/0$), which can be solved by Taylor expansion to give the acoustic natural angular frequency of the duct.
From a classical mechanics standpoint, $G/2$ is the Lagrangian, which is the difference between the kinetic and potential energies.
We are interested in the eigenvalue with $\sigma_i>0$, so the sign of the right-hand side is positive. Hence the acoustic potential energy is smaller (larger) than the acoustic kinetic energy when the imaginary part of the Rayeligh integral~\eqref{eq:llqqkgj4893} is negative (positive).
%
%
%
%
Equations~\eqref{eq:ffrwu52478} and \eqref{eq:llqqkgj4893} represent the Rayleigh criterion in the frequency domain, which states that if the heat release rate is in phase with the pressure, the growth rate is maximized. We also gain an extra piece of information with respect to the time domain analysis: if the heat release rate is in quadrature with the pressure, the frequency is maximized. The latter is the mathematical formalization and generalization of a statement made in~\citet{Rayleigh1878} regarding the pitch of the oscillation. The integral formulae are showcased analytically in \S2 of the supplementary material. 
%
%
%
%
%
%
%
%
%
\subsection{First variation without Lagrange multipliers}\label{sec:ffflkpp000003}
We follow the same procedure as that of \S\ref{sec:firstvariation_time} with the modifications needed to switch from the time domain to the frequency domain. We perturb the heat release rate by an arbitrary function
$\hat{\dot{Q}}(x) \rightarrow \hat{\dot{Q}}(x) + \epsilon\hat{\dot{Q}}_p(x)$ 
where $\hat{\dot{Q}}_p(x)$ is the test function in the frequency domain. 
The objective is to find the first variation of the eigenvalue, $\delta\sigma$,  in the vicinity of $\hat{p}$ and $\hat{\dot{Q}}$. 
%
The eigenvalue functional derivative we wish to calculate is the kernel of the eigenvalue first variation 
\begin{align}
\delta\sigma=\epsilon\int_0^L 
{
\left(\frac{\delta_{\hat{\dot{Q}}}\sigma}{\delta\hat{\dot{Q}}(x')}
+
\frac{\delta_{\hat{p}}\sigma}{\delta\hat{p}(x')}\frac{d\hat{p}^{\lmf{*}}}{d\hat{\dot{Q}}(x')^{\lmf{*}}}
\right)^*}
\hat{\dot{Q}}_p(x')\;dx'.
\end{align}
A similar physical explanation to that given in \S\ref{sec:firstvariation_time} can be given here in terms of eigenfunctions. The two eigenvalue functional derivatives are
\begin{align}
\frac{\delta_{\hat{\dot{Q}}} \sigma}{\delta\hat{\dot{Q}}} & = \lmf{\frac{1}{C^*}\left(\frac{\gamma-1}{\gamma\bar{p}}\right)}\hat{p}^{\lmf{*}}, \quad\quad\quad\quad\quad\quad \frac{\delta_{\hat{p}} \sigma}{\delta \hat{p}}  = \lmf{\frac{1}{C^*}\left(\frac{\gamma-1}{\gamma\bar{p}}\right)}\hat{\dot{Q}}^{\lmf{*}}, 
\end{align}
%
%
%
%
%
%
%
%
whence the eigenvalue functional derivative is 
\begin{align}
\frac{\delta\sigma}{\delta\hat{\dot{Q}}} \coloneqq \left(\frac{\delta_{\hat{\dot{Q}}}\sigma}{\delta\hat{\dot{Q}}(x')}
+
\frac{\delta_{\hat{p}}\sigma}{\delta\hat{p}(x')}\frac{d\hat{p}^{\lmf{*}}}{d\hat{\dot{Q}}(x')^{\lmf{*}}}\right)= \frac{1}{C^{\lmf{*}}}\left(\frac{\gamma-1}{\gamma\bar{p}}\right)\left(\hat{p} + \frac{d\hat{p}}{d\hat{\dot{Q}}}\hat{\dot{Q}}\right)^{\lmf{*}}, \label{eq:llkjmnmmf40}
\end{align} 
such that the eigenvalue first variation is provided by $\delta\sigma
= 
\langle  \lmf{({\delta\sigma}/{\delta\hat{\dot{Q}}})}, \epsilon\hat{\dot{Q}}_p\rangle_V$.
\subsection{First variation with Lagrange multipliers}\label{eq:FirstvariationwithLagrangemultipliers}
The objective of this section is the same as that of \S\ref{sec:firstvariation_time} but in the frequency domain. We seek the first variation of the eigenvalue expressed as a functional of the Lagrangian multiplier. 
The problem is cast as a constrained optimization
problem: we wish to find the eigenvalue functional derivative ${\delta \sigma}/{\lmm{\delta\hat{\dot{{Q}}}}}$ subject to~\eqref{eq:lop43}, \eqref{eq:lop432}. 
We define a Lagrangian  
\begin{align}
\mathcal{L}(\sigma, \hat{u}, \hat{p}, \hat{u}^+, \hat{p}^+, \hat{\dot{Q}})\coloneqq \sigma - \left\langle\hat{u}^+, \eqref{eq:lop43}\right\rangle_V - \left\langle\hat{p}^+, \eqref{eq:lop432}\right\rangle_V. 
\end{align}
%
%
After integrating by parts, the first variation of the Lagrangian reads  
%
\begin{align}
\delta\mathcal{L}&=\delta\sigma\left(1-\int_0^L\left(\bar{\rho}\hat{u}^{+*}\hat{u}+\frac{1}{\gamma\bar{p}}\hat{p}^{+*}\hat{p} \lmm{\;-\;} \lmm{\frac{\gamma-1}{\gamma\bar{p}}\frac{\partial\hat{\dot{{Q}}}}{\partial\sigma}\hat{p}^{+*}}\right)\;dx\right) \ldots\nonumber\\
 &- [{\hat{u}^{+*}\delta\hat{p}}]_0^L -
 [{ \hat{p}^{+*}\delta\hat{u}}]_0^L + \left\langle
-\sigma\bar{\rho}\hat{u}^{+} + \frac{\partial \hat{p}^{+}}{\partial x} + \lmm{\frac{\gamma-1}{\gamma\bar{p}}\left(\frac{\partial\hat{\dot{{Q}}}}{\partial\hat{u}}\right)^*\hat{p}^{+}} , \delta\hat{u}\right\rangle_V\ldots\nonumber\\
&+\left\langle-\sigma\frac{\hat{p}^{+}}{\gamma\bar{p}} + \frac{\partial \hat{u}^{+}}{\partial x} + \lmm{\frac{\gamma-1}{\gamma\bar{p}}\left(\frac{\partial\hat{\dot{{Q}}}}{\partial\hat{p}}\right)^*\hat{p}^{+}}, \delta\hat{p}\right\rangle_V
 +\epsilon\frac{\gamma-1}{\gamma\bar{p}}\left\langle\lmb{\hat{p}^{+}, \hat{\dot{{Q}}}_p}\right\rangle_V. \label{e38203208023409432}
\end{align}
We define the adjoint eigenvalue problem  
\begin{align}\label{eq:for3f2j43ij05}
 -\sigma \bar{\rho}\hat{u}^{+}+\frac{\partial \hat{p}^{+}}{\partial x}+\lmm{\frac{\gamma-1}{\gamma\bar{p}}\left(\frac{\partial\hat{\dot{{Q}}}}{\partial\hat{u}}\right)^*\hat{p}^{+}}&=0, \\ \label{eq:for3f2j43ij052}
 -\sigma \frac{\hat{p}^{+}}{\gamma\bar{p}}+\frac{\partial \hat{u}^{+}}{\partial x} + \lmm{\frac{\gamma-1}{\gamma\bar{p}}\left(\frac{\partial\hat{\dot{{Q}}}}{\partial\hat{p}}\right)^*\hat{p}^{+}}&=0, 
\end{align}
with adjoint boundary conditions $[\hat{u}^{+*}\delta\hat{p}]_0^L =0$ and $[\hat{p}^{+*}\delta\hat{u}]_0^L =0$.   
Hence, the first variation of the eigenvalue can be calculated as $\delta\sigma=\delta\mathcal{L}
=(1/{D}) (\gamma-1)/(\gamma\bar{p}){\langle \hat{p}^{+}, \epsilon\hat{\dot{Q}}_p\rangle_V}$, where $D \coloneqq \int_0^L(\bar{\rho}\hat{u}^{+*}\hat{u}+\frac{1}{\gamma\bar{p}}\hat{p}^{+*}\hat{p}\lmm{\;-\;} \lmm{\frac{\gamma-1}{\gamma\bar{p}}\hat{p}^{+*}\frac{\partial\hat{\dot{{Q}}}}{\partial\sigma}})\;dx$. 
Finally, the eigenvalue functional derivative is 
\begin{align}
\frac{\delta\sigma}{\delta\hat{\dot{Q}}} = \frac{1}{D^{\lmf{*}}}\left(\frac{\gamma-1}{\gamma\bar{p}}\right)\hat{p}^{+}.\label{eq:llkjmnmmf4}
\end{align} 
\subsection{Relation between first variations with and without Lagrange multipliers}
%
%
%
%
%
The relations~\eqref{eq:llkjmnmmf40} and~\eqref{eq:llkjmnmmf4} imply that   
\begin{equation}\label{eq:lldo32}
\hat{p}^{+*}  = \frac{D}{C}\left(
\hat{p}
+ \frac{d\hat{p}}{d\hat{\dot{Q}}}\hat{\dot{Q}}\right). 
\end{equation}
This is a key equation in the frequency domain.
It is the link between the Lagrange multiplier (adjoint variable) and the system's observables. 
%
Physically, the adjoint eigenfunction is the sum of two terms multiplied by a constant $D/C$. Starting from the first term on the right-hand side they are: 
(i) The  acoustic pressure at the location where the perturbation is applied (acoustic sensitivity), and 
(ii) the acoustic pressure sensitivity at the main source spatial support weighted by the main heat release rate (thermal sensitivity). Relation~\eqref{eq:lldo32} shows how to express the first variation of the Rayleigh criterion in the frequency domain in a non-self-adjoint system ($\hat{\dot{Q}}\neq0$); as in~\S\ref{sec:fkfkfkfkfk2111116}, non-self-adjointness manifests itself via the thermal sensitivity $(d \hat{p}/d\hat{\dot{Q}})\hat{\dot{Q}}$, which is a nonlocal effect.   
For a localized source, similarly to~\S\ref{sec:fkfkfkfkfk2111116},  
${\hat{p}_p^{+*}}  = ({D}/{C}) 
(
\hat{p}_p
+ {d\hat{p}_f}\hat{{Q}}_f/({\epsilon\hat{{Q}}_p}))$.

 \lmr{Relation~\eqref{eq:lldo32} can be used in two ways. On the one hand, we can obtain an approximation of the adjoint variable by interpolating experimental measurements of $\hat{p}+(d\hat{p}/d\hat{\dot{Q}})\hat{\dot{Q}}$. On the other hand, we can cheaply estimate the sensitivity function from a low-order model as ${\hat{p}^{+*}} - (D/C)\hat{p}$. This indicates which areas of the flow should be sampled to reconstruct an accurate sensitivity function. In other words, this relation gives a prior to guide the measurements in experiments and sampling in high-fidelity simulations. 
}
\subsection{The adjoint Rayleigh criterion}\label{eq:fijij8hur3fuihfhuifr}
The adjoint Rayleigh criterion answers the question: ``Under which conditions is the first variation of the eigenvalue positive, negative, or zero, when the heat source is perturbed?". This was numerically investigated in~\citet{Magri2013,Magri2013c,Juniper2018_prf}. Here, we formalize the adjoint Rayleigh criterion following the argument of~\citet[Appendix C,][]{Magri2019_amr}. %
The factor $D$ can be set to a real number~\citep[e.g., Eq.~(28), ][]{Aguilar2017}, so we set $D=(\gamma-1)/(\gamma\bar{p})$ in this section without loss of generality. 
For localized sources, the first variation of the growth rate is 
$\delta\sigma_r  =\lvert\hat{p}_p^{*+}\lvert \lvert\hat{\dot{Q}}_p\lvert\cos(\Delta\theta)$ where $\Delta\theta$ is the difference between the arguments of the heat release rate and \lmr{adjoint} pressure. 
%
%
%
%
%
%
%
This means that  $\delta\sigma_r$ is maximum when $\Delta\theta=\pm 2(k-1)\pi$, where $k$ is a positive integer, i.e., when a perturbation to the heat release rate is in phase with the adjoint pressure, the system is most destabilized. 
Second, the first variation of the growth rate, $\delta\sigma_r$, is minimum when $\Delta\theta=\pm (2k-1)\pi$. i.e., when a perturbation to the heat release rate is in antiphase with the adjoint pressure, the system is most stabilized. %
Third, the first variation of the growth rate $\delta\sigma_r$ is zero when $\Delta\theta=\pm (2k+1)\pi/2$., i.e., when a perturbation to the heat release rate is in quadrature with the adjoint pressure, the system's stability is unaffected. 
The first variation of the angular frequency is  
$\delta\sigma_i  = \lvert\hat{p}_p^{*+}\lvert \lvert\hat{\dot{Q}}_p\lvert\sin(\Delta\theta)$.
In contrast to the growth rate, $\delta\sigma_i$ is maximum when the adjoint pressure is in quadrature with the heat release rate. Note that we could have used the adjoint without the complex conjugation in~\eqref{eq:innerproductddd}. The results discussed would still be valid by taking into account that, without complex conjugate, the adjoint eigenvector rotates in time as $\exp(-\sigma t)$, which is to say that the angular frequency is negative.
\section{Conclusions}\label{sec:conclusions}
Thermoacoustic instabilities are a major challenge for the gas turbine and rocket motor industries.
The Rayleigh criterion is widely used to establish whether or not a thermoacoustic system is unstable. 
%
One result of this paper is to formalize and interpret the Rayleigh criterion in the frequency domain (\S\ref{eq:lldafio30}).
This quantifies a phenomenon that was qualitatively described by \citet{Rayleigh1878}: if the heat release rate is in quadrature with the pressure, the frequency is maximized. 
Another result is to derive the first variation of the Rayleigh criterion in both the time and the frequency domains when an extra heat source is applied.
In each domain, we perform analysis both with and without Lagrange multipliers (adjoint variables).
The change in the acoustic energy (time domain) or the change in the growth rate/frequency (frequency domain) is given by integrals of the heat release rate perturbation multiplied by the adjoint pressure.
Further, we derive exact relationships (\eqref{eq:ffe3344d}, \eqref{eq:lldo32}) between the adjoint pressure and the observable quantities. 
These relationships show how to express the first variation of the Rayleigh criterion in non-self-adjoint systems.
%
We show that the adjoint pressure, which is physically interpreted as the functional derivative of the acoustic energy (or eigenvalue) with respect to a perturbation to the heat source, consists of two terms: (i) the pressure at the location at which the perturbation is applied, which is the acoustic sensitivity; and (ii) the acoustic pressure sensitivity, weighted by the main heat release rate, at the main source location, which is the thermal sensitivity. 
The superposition of these two sensitivities gives the thermoacoustic sensitivity.
Non-self-adjointness manifests itself as a nonlocal effect (the thermal sensitivity). 
%
%
Finally, an adjoint Rayleigh criterion is proposed. This criterion provides, in a compact form, precise conditions on the effect that perturbations to the heat source have on the stability of a thermoacoustic system. 

The mathematical relations of this paper can be used as diagnostic tools for the interpretation of experimental results, and as methods to compute sensitivities directly from measurable experimental quantities, which is useful for optimal design by experimental sensitivity analysis. \lmr{In future work, the approach of this paper can be applied with virtually no modification to the three-dimensional acoustic equations with a mean flow.}
%



%
%
%
%

%
%
\bibliographystyle{jfm}
    \bibliography{MyCollection}
    
\end{document}